\PassOptionsToPackage{unicode}{hyperref}
\PassOptionsToPackage{hyphens}{url}
\documentclass[
]{article}
\usepackage{amsmath,amssymb}
\usepackage{lmodern}
\usepackage{iftex}
\ifPDFTeX
  \usepackage[T1]{fontenc}
  \usepackage[utf8]{inputenc}
  \usepackage{textcomp} 
\else 
  \usepackage{unicode-math}
  \defaultfontfeatures{Scale=MatchLowercase}
  \defaultfontfeatures[\rmfamily]{Ligatures=TeX,Scale=1}
\fi
\IfFileExists{upquote.sty}{\usepackage{upquote}}{}
\IfFileExists{microtype.sty}{
  \usepackage[]{microtype}
  \UseMicrotypeSet[protrusion]{basicmath} 
}{}
\makeatletter
\@ifundefined{KOMAClassName}{
  \IfFileExists{parskip.sty}{%
    \usepackage{parskip}
  }{
    \setlength{\parindent}{0pt}
    \setlength{\parskip}{6pt plus 2pt minus 1pt}}
}{
  \KOMAoptions{parskip=half}}
\makeatother
\usepackage{xcolor}
\usepackage[margin=1in]{geometry}
\usepackage{graphicx}
\makeatletter
\def\maxwidth{\ifdim\Gin@nat@width>\linewidth\linewidth\else\Gin@nat@width\fi}
\def\maxheight{\ifdim\Gin@nat@height>\textheight\textheight\else\Gin@nat@height\fi}
\makeatother
\setkeys{Gin}{width=\maxwidth,height=\maxheight,keepaspectratio}
\makeatletter
\def\fps@figure{htbp}
\makeatother
\providecommand{\tightlist}{%
  \setlength{\itemsep}{0pt}\setlength{\parskip}{0pt}}
\usepackage{multirow}
\usepackage{multicol}
\usepackage{colortbl}
\usepackage{hhline}
\newlength\Oldarrayrulewidth
\newlength\Oldtabcolsep
\usepackage{longtable}
\usepackage{array}
\usepackage{hyperref}
\usepackage{float}
\usepackage{wrapfig}
\usepackage[numbers]{natbib}
\usepackage{authblk}
\ifLuaTeX
  \usepackage{selnolig}  
\fi
\IfFileExists{bookmark.sty}{\usepackage{bookmark}}{\usepackage{hyperref}}
\IfFileExists{xurl.sty}{\usepackage{xurl}}{} 
\hypersetup{
  pdftitle={Review of EMA EPARs where nonproportional hazards were identified},
  pdfauthor={Tobias Fellinger, Norbert Benda, Tim Friede, Harald Heinzl, Andrew Hooker, Franz König, Tim Mathes, Martin Posch, Susanne Urach, Florian Klinglmueller},
  hidelinks,
  pdfcreator={LaTeX via pandoc}}

\title{A Review of EMA Public Assessment Reports where Non-Proportional Hazards were
Identified}
\author[1]{Florian Klinglmueller}
\author[2]{Norbert Benda}
\author[3]{Tim Friede}
\author[1]{Tobias Fellinger}
\author[4]{Harald Heinzl}
\author[5]{Andrew Hooker}
\author[6]{Franz König}
\author[3]{Tim Mathes}
\author[6]{Martin Posch}
\author[1]{Florian Stampfer}
\author[1]{Susanne Urach}

\affil[1]{Austrian Agency for Health and Food Safety (AGES), Vienna, Austria}
\affil[2]{Federal Institute for Drugs and Medical Devices, Bonn, Germany}
\affil[3]{Department of Medical Statistics, University Medical Center Göttingen, Göttingen, Germany}
\affil[4]{Center for Medical Data Science, Section of Clinical Biometrics, Medical University of Vienna, Vienna, Austria}
\affil[5]{Department of Pharmacy, Uppsala University, Uppsala, Sweden}
\affil[6]{Center for Medical Data Science, Section of Medical Statistics, Medical University of Vienna, Vienna, Austria}
\date{12 6 2024}

\begin{document}
\maketitle

\hypertarget{abstract}{%
\section{Abstract}\label{abstract}}

\textbf{Objective:} Identify EMA marketing authorization procedures
where non-proportional hazards (NPH) were of concern in the efficacy
assessment in order to define relevant parameter ranges for a simulation
study, derive case studies and regulatory recommendations.

\textbf{Introduction:} While well-established methods for time-to-event
data are available when the proportional hazards assumption holds, there
is no consensus on the best approach under non-proportional hazards.
However, a wide range of parametric and non-parametric methods for
testing and estimation in this scenario have been proposed. In this
review we identified EMA marketing authorization procedures where
non-proportional hazards were raised as a potential issue in the
risk-benefit assessment and extract relevant information on trial result
reported in the corresponding European Assessment Reports (EPARs).

\textbf{Inclusion criteria:} We included marketing authorization
procedures granted before March 1st 2022. Procedures were limited to
initial authorizations as well as variations e.g.~extending the use to
other therapeutic areas, for which the European Public Assessment Report
(EPAR) reports results from at least one pivotal comparative trial and
where a method that accounts for deviations from the proportional
hazards assumption was used for the analysis of a primary (or
key-secondary) endpoint in at least one pivotal trial. Procedures, where
the use of a method to address non-proportional hazards was limited to
sensitivity analyses and no notable results were reported were excluded.

\textbf{Methods:} EPARs available in the database at paediatricdata.eu
were searched for paragraphs matching a predefined list of keywords
related to the assumption of non-proportional hazards. Results were
screened to exclude matches unrelated to NPH issues or descriptions of
routine sensitivity analysis. For selected matches full-text of EPARs
were obtained and information on to the suspected cause of NPH, the
methods used to address NPH, and results from statistical analyses were
extracted.

\textbf{Results} We identified 16 Marketing authorization procedures,
with EPARs reporting results on a total of 18 trials. Most procedures
covered the authorization of treatments from the oncology domain. For
the majority of trials NPH issues were related to a suspected delayed
treatment effect, or different treatment effects in known subgroups.
Issues related to censoring, or treatment switching were also
identified. For most of the trials the primary analysis was performed
using conventional methods assuming proportional hazards, even if NPH
was anticipated. Differential treatment effects were addressed using
stratification and delayed treatment effect considered for sample size
planning. Even though, not considered in the primary analysis, some
procedures reported extensive sensitivity analyses and model diagnostics
evaluating the proportional hazards assumption. For a few procedures
methods addressing NPH (e.g.~weighted log-rank tests) were used in the
primary analysis. We extracted estimates of the median survival, hazard
ratios, and time of survival curve separation. In addition, we digitized
the KM curves to reconstruct close to individual patient level data.
Extracted outcomes served as the basis for a simulation study
\citep{Klinglmuller2023} of methods for time to event analysis under
NPH.

\hypertarget{introduction}{%
\section{Introduction}\label{introduction}}

The main objective of this review was to identify EMA marketing
authorization procedures where non-proportional hazards were identified
as a potential issue in the risk-benefit assessment of the corresponding
medicinal product. Based on this review, we will define relevant
parameter ranges for distributional scenarios for a simulation study, to
derive case studies in order to illustrate the findings of the
simulation study, and derive regulatory recommendations.

This review focuses on marketing procedures where non-proportional
hazards were identified as a potential issue. The main question of this
review addresses: for which past marketing authorization procedures were
non-proportional hazards identified as a potential issue in the
assessment. Thus, we investigated which methods were used to address
potential violations of the proportional hazards assumption and which
reasons were suspected as cause for violations of the proportional
hazards assumption. In addition, we extracted aggregate data on clinical
trial endpoints and reconstructed close to individual patient level data
from KM curves in order to define relevant parameter ranges for a
simulation study and derive case studies for illustration.

\hypertarget{eligibility-criteria}{%
\section{Eligibility criteria}\label{eligibility-criteria}}

To identify marketing procedures where non-proportional hazards were
identified as a potential issue, the following eligibility criteria were
defined in the protocol:

\hypertarget{inclusion-criteria}{%
\subsubsection{Inclusion criteria}\label{inclusion-criteria}}

\begin{enumerate}
\def\labelenumi{\arabic{enumi}.}
\tightlist
\item
  Marketing authorization procedures with a positive opinion (granted
  before March 1st 2022) where non-proportional hazards were identified
  as a potential issue during the assessment of efficacy for the
  corresponding medicinal product.
\item
  Initial authorization or variation e.g.~extending the authorised use
  to another therapeutic area.
\item
  Procedures for which the EPAR reports results on at least one pivotal
  comparative trial.
\item
  A method that accounts for deviations from the proportional hazards
  assumption was used for the analysis of a primary (or key-secondary)
  endpoint in at least one pivotal confirmatory trial and where
  corresponding concerns are reflected in the EPAR.
\end{enumerate}

\hypertarget{exclusion-criteria}{%
\subsubsection{Exclusion criteria}\label{exclusion-criteria}}

\begin{enumerate}
\def\labelenumi{\arabic{enumi}.}
\tightlist
\item
  Marketing authorization procedures currently under review.
\item
  Marketing authorization procedures withdrawn by the Applicant or with
  a negative opinion.
\item
  Procedures for which the EPAR does not report results on at least one
  pivotal comparative trial.
\item
  The use of a method to address non-proportional hazards was limited to
  sensitivity analyses to evaluate a potential deviation from the
  proportional hazards assumption and where no concern was raised in the
  EPAR.
\end{enumerate}

The eligibility criteria are formulated in complementary pairs of in-
and exclusion criteria. The first and second set of criteria (in- and
exclusion criteria 1 and 2) limit the eligible procedures to completed
procedures with a positive opinion. Completed procedures with a negative
opinion or withdrawn by the Applicant were excluded as the elected
database at paediatricdata.eu does not comprehensively cover
corresponding procedures. The third set of criteria (in- and exclusion
criteria 3) limits procedures, for which the EPAR reports results on a
pivotal comparative trial. Procedures e.g.~reporting results only from
supportive studies or single arm trial, were not considered relevant for
the review. Finally, the fourth set of criteria (in- and exclusion
criteria 4) limits the included procedures to those that report at least
some relevant and notable results from analyses addressing
non-proportional hazards.

\hypertarget{methods}{%
\section{Methods}\label{methods}}

\hypertarget{search-strategy}{%
\subsubsection{Search strategy}\label{search-strategy}}

A comprehensive search of EMA Public Assessment Reports (EPARs) was
conducted in the electronic database \url{https://paediatricdata.eu}.
Search terms included terms such as ``non-proportional hazards'' in
different types of spelling and related prespecified search terms. Due
to limitations of the database provided at
\url{https://paediatricdata.eu} search was restricted to procedures with
a positive opinion.

An initial limited search of paediatricdata.eu was undertaken to
evaluate the feasibility of a search strategy based on full-text queries
of EPARs. The final search strategy, including all search terms and
conditions for selection was adapted from the systematic review of the
statistical literature \citep{Bardo2024}. However, only rules
applicable to items provided by the respective databases were included.

Search results for individual search terms were exported as
spreadsheets, including for each result at least identifiers for
procedure, active substance, matching paragraph text, and source file.
Duplicates were removed and remaining EPARs were filtered to match
predefined combination rules of search terms. The list of search terms
and combination rules is provided in Figure \ref{fig:unnamed-chunk-5} in the Appendix.

The search interface for the \url{https://paediatricdata.eu} database
can be found at
\url{https://paediatricdata.eu/shiny/users/ralfherold/emaepars/}. The
search for Terms 1 to 9 was performed on March 9 2022, search for Terms
10 to 20 on March 20 2022 (following an extension of search terms for
the literature review requested by EMA).

\hypertarget{selection-of-marketing-authorization-procedures}{%
\subsubsection{Selection of marketing authorization
procedures}\label{selection-of-marketing-authorization-procedures}}

Matching paragraph text from EPARs were screened for assessment against
the inclusion criteria for the review. Screening was be performed using
a screening form developed by the reviewers. For this purpose, a small
number of search results were screened in a pilot test by AGES reviewers
to evaluate the usability of the form. Subsequently, all paragraphs
identified by the database search were grouped by procedure, and
screening was performed in pairs of independent reviewers from different
institutions. This was implemented to minimize inadvertent exchange
between reviewers and ensure independence of assessment. In case of
disagreement between reviewers regarding inclusion, the procedure was
included for data extraction, anyway. Only procedures where reviewers
agreed on exclusion were excluded at this stage.

For the selected procedures the EPAR was retrieved in full. The full
text of selected procedures were assessed against the eligibility
criteria by two independent reviewers. It was ensured that only
reviewers not responsible for the screening of a particular procedure
were assigned to assess eligibility based on the full-text reports.
Procedures with conflicting screening results were preferably assigned
to experts not included in the screening step. Any disagreements that
arose between the reviewers at the full-text screening stage were
resolved by a third reviewer. The results of the search and the study
inclusion process reported in below and presented in a Preferred
Reporting Items for Systematic Reviews and Meta-analyses extension for
scoping review (PRISMA-ScR) flow diagram \citep{Tricco2018} in Figure
1.

\hypertarget{data-extraction}{%
\subsubsection{Data Extraction}\label{data-extraction}}

Reviewers assessing the eligibility of EPARs based on full-text review
(see above) extracted data on trials reported in EPARs using a data
extraction form developed by the authors (See Supplement 2 for an
example extraction form). The data extracted included specific details
about the procedure, product, indication, suspected potential reasons
for NPH, methods to address NPH, as well as key aggregate level study
results.

Completed extraction forms were merged for each procedure. Duplicate
text and images were removed. Extracted descriptions of statistical
methods and results were summarized and copy-edited to improve
legibility. Included screenshots of tables and figures with poor quality
were replaced with higher resolution screenshots from the full-text
report. Form contents were extracted and concatenated in spreadsheets.
For included procedures a summary report including the extracted
information (removing instructions, inclusion criteria, special interest
flags, and comments) was prepared (see Supplement 1).

\hypertarget{data-analysis-and-presentation}{%
\subsubsection{Data Analysis and
Presentation}\label{data-analysis-and-presentation}}

In general, data - extracted from EPARs are summarised using descriptive
statistics. We give an overview on the number of procedures identified.
Absolute and relative frequencies of broad and specific indication
areas, types of and suspected cause for NPH concerns, endpoints, and
methods to address NPH. We provide an overview for a range of relevant
outcome measures including hazard ratios and median survival. For trials
where a delayed treatment effect was suspected we report the time of
separation (or crossing) of survival curves. For trials, where a
differential treatment effect was suspected within subgroups we report
treatment effect measures per subgroup.

\hypertarget{results}{%
\section{Results}\label{results}}

\hypertarget{study-selection}{%
\subsubsection{Study selection}\label{study-selection}}

At the time the final search was performed (March 16, 2022) the database
at paediatricdata.eu held records on 1561 procedures and more than 4500
(exact number was not saved at the time of search) full-text reports
referring to procedures with an authorization date up to February 21,
2022 (corresponding to the February Meeting of the European Medicines
Agencies, Committee for Medicinal Products for Human Use). The combined
keyword search return 133 paragraphs from 92 documents in the database.
Returned paragraphs were grouped by procedure and screened by pairs of
independent reviewers (see above). Following this screening step 28 out
of 92 procedures were selected to be included by at least one reviewer.
One additional procedure \emph{EMEA/H/C/003874/0000} (Invented Name:
Onzeald, Active Substance: etirinotecan pegol), was referred to us by
the EMA members in the Steering Committee. This procedure did not
receive a positive opinion and was therefore selected to be included as
a special case. Consequently, a total of 29 procedures were included for
full-text review.

\begin{figure}
\includegraphics[width=6in]{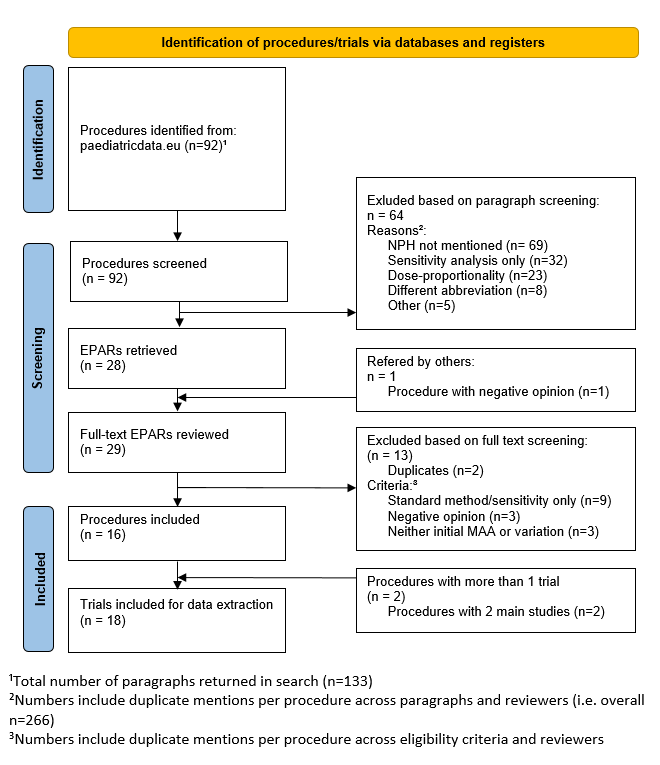} \caption{PRISMA 2020 flow diagram - identified and included procedures for the review of EPARs.}\label{fig:prism}
\end{figure}

Of the 29 procedures included for full-text review, reviewers selected
16 procedures for data-extraction and inclusion in the review. Of the
remaining 13 procedures two duplicate procedures were removed and 11
procedures were either excluded unanimously or in case of disagreement
following re-assessment by a third reviewer. For two of the included
procedures the EPAR reported results from two main studies.
Consequently, a data from a total of 18 trials were extracted for the
review. Figure \ref{fig:prism} provides the PRISM Flowchart for the inclusion process,
giving also some details on the type and frequency of reasons for
exclusion.

The majority of procedures were excluded due to inclusion criterion 4
and corresponding exclusion criterion 4. \emph{I.e.:} Include if: ``A
method that accounts for deviations from the proportional hazards
assumption was used for the analysis of a primary (or key-secondary)
endpoint in at least one pivotal confirmatory trial and where
corresponding concerns are reflected in the EPAR'', or exclude if: ``The
use of a method to address non-proportional hazards was limited to
sensitivity analyses to evaluate a potential deviation from the
proportional hazards assumption and where no concern was raised in the
EPAR''. As it turned out, these criteria were not consistently applied
in case NPH issues were apparent and discussed in the EPAR, but the
primary analysis was performed using a model assuming proportional
hazards (e.g.~stratified Cox Regression). To resolve corresponding
disagreements, procedures were re-assessed and included if some results
of sensitivity analyses or model diagnostics with considerable results
were reported, but excluded if NPH was not discussed or sensitivity
analyses and model diagnostics did not indicate a violation of NPH.

\hypertarget{included-procedures-by-indication-procedure-type-and-active-substance}{%
\section{Included procedures by indication, procedure type, and active
substance}\label{included-procedures-by-indication-procedure-type-and-active-substance}}

We grouped the indications sought by the different procedures included
in the review by broad indication areas classes. Table 1 provides the
number of procedures per indication area. The large majority of
procedures are from the oncology domain. For the other areas -
cardiovascular disease, conscious sedation, influenza (human), and
multiple sclerosis - only a single procedure was identified.

\global\setlength{\Oldarrayrulewidth}{\arrayrulewidth}

\global\setlength{\Oldtabcolsep}{\tabcolsep}

\setlength{\tabcolsep}{0pt}

\renewcommand*{\arraystretch}{1.5}

\providecommand{\ascline}[3]{\noalign{\global\arrayrulewidth #1}\arrayrulecolor[HTML]{#2}\cline{#3}}

\begin{longtable}[c]{|p{1.84in}|p{1.35in}}

\caption{Number\ of\ procedures\ by\ indication\ area}\\

\ascline{1.5pt}{666666}{1-2}

\multicolumn{1}{>{\raggedright}m{\dimexpr 1.84in+0\tabcolsep}}{\textcolor[HTML]{000000}{\fontsize{11}{11}\selectfont{Indication\ Area}}} & \multicolumn{1}{>{\raggedleft}m{\dimexpr 1.35in+0\tabcolsep}}{\textcolor[HTML]{000000}{\fontsize{11}{11}\selectfont{\#\ of\ procedures}}} \\

\ascline{1.5pt}{666666}{1-2}\endhead

\multicolumn{1}{>{\raggedright}m{\dimexpr 1.84in+0\tabcolsep}}{\textcolor[HTML]{000000}{\fontsize{11}{11}\selectfont{oncology}}} & \multicolumn{1}{>{\raggedleft}m{\dimexpr 1.35in+0\tabcolsep}}{\textcolor[HTML]{000000}{\fontsize{11}{11}\selectfont{12}}} \\

\multicolumn{1}{>{\raggedright}m{\dimexpr 1.84in+0\tabcolsep}}{\textcolor[HTML]{000000}{\fontsize{11}{11}\selectfont{cardiovascular\ disease}}} & \multicolumn{1}{>{\raggedleft}m{\dimexpr 1.35in+0\tabcolsep}}{\textcolor[HTML]{000000}{\fontsize{11}{11}\selectfont{1}}} \\

\multicolumn{1}{>{\raggedright}m{\dimexpr 1.84in+0\tabcolsep}}{\textcolor[HTML]{000000}{\fontsize{11}{11}\selectfont{concious\ sedation}}} & \multicolumn{1}{>{\raggedleft}m{\dimexpr 1.35in+0\tabcolsep}}{\textcolor[HTML]{000000}{\fontsize{11}{11}\selectfont{1}}} \\

\multicolumn{1}{>{\raggedright}m{\dimexpr 1.84in+0\tabcolsep}}{\textcolor[HTML]{000000}{\fontsize{11}{11}\selectfont{influenza\ (human)}}} & \multicolumn{1}{>{\raggedleft}m{\dimexpr 1.35in+0\tabcolsep}}{\textcolor[HTML]{000000}{\fontsize{11}{11}\selectfont{1}}} \\

\multicolumn{1}{>{\raggedright}m{\dimexpr 1.84in+0\tabcolsep}}{\textcolor[HTML]{000000}{\fontsize{11}{11}\selectfont{multiple\ sclerosis}}} & \multicolumn{1}{>{\raggedleft}m{\dimexpr 1.35in+0\tabcolsep}}{\textcolor[HTML]{000000}{\fontsize{11}{11}\selectfont{1}}} \\

\ascline{1.5pt}{666666}{1-2}

\end{longtable}

\arrayrulecolor[HTML]{000000}

\global\setlength{\arrayrulewidth}{\Oldarrayrulewidth}

\global\setlength{\tabcolsep}{\Oldtabcolsep}

\renewcommand*{\arraystretch}{1}

\hypertarget{active-substances-investigated-in-the-included-procedures}{%
\subsubsection{Active substances investigated in the included
procedures}\label{active-substances-investigated-in-the-included-procedures}}

Most of the procedures included (n=12) represent variations, \emph{e.g.}
extending the authorized use of a product to a new disease area or
population. Not all of the procedures concern different active
substances. For pembrolizumab (n=5), ipilimumab (n=2), and nivolumab
(n=2) more than one procedure was included. Table 2 presents the number
of procedures per active substance and procedure type. The majority of
compounds represent monoclonal antibodies (i.e.~all compounds ending
with \emph{-mab}), a type of immunotherapy used in cancer treatment.
Specifically, the identified compounds belong to the class of checkpoint
inhibitors, which block specific proteins that stop the immune system
attacking cancer cells \citep{Gong2018}. The efficacy of corresponding
therapies therefore depends on the amount of the corresponding protein
target on a patients cancer cells. Consequently, a differential
treatment effect in known biomarker defined subgroups, resulting in a
violation of the proportional hazards assumption was anticipated for
these therapies \citep{Havel2019}. In addition, for the corresponding
immunotherapies issues affecting the proportional hazards assumption
like delayed treatment effects, and long-term survivors have been
reported \citep{Mick2015}.

\global\setlength{\Oldarrayrulewidth}{\arrayrulewidth}

\global\setlength{\Oldtabcolsep}{\tabcolsep}

\setlength{\tabcolsep}{0pt}

\renewcommand*{\arraystretch}{1.5}

\providecommand{\ascline}[3]{\noalign{\global\arrayrulewidth #1}\arrayrulecolor[HTML]{#2}\cline{#3}}

\begin{longtable}[c]{|p{1.52in}|p{1.54in}|p{1.35in}}

\caption{Number\ of\ procedures\ by\ active\ substance\ and\ type\ of\ authorization\ procedure}\\

\ascline{1.5pt}{666666}{1-3}

\multicolumn{1}{>{\raggedright}m{\dimexpr 1.52in+0\tabcolsep}}{\textcolor[HTML]{000000}{\fontsize{11}{11}\selectfont{Active\ Substance}}} & \multicolumn{1}{>{\raggedright}m{\dimexpr 1.54in+0\tabcolsep}}{\textcolor[HTML]{000000}{\fontsize{11}{11}\selectfont{Type\ of\ Procedure}}} & \multicolumn{1}{>{\raggedleft}m{\dimexpr 1.35in+0\tabcolsep}}{\textcolor[HTML]{000000}{\fontsize{11}{11}\selectfont{\#\ of\ procedures}}} \\

\ascline{1.5pt}{666666}{1-3}\endhead

\multicolumn{1}{>{\raggedright}m{\dimexpr 1.52in+0\tabcolsep}}{\textcolor[HTML]{000000}{\fontsize{11}{11}\selectfont{pembrolizumab}}} & \multicolumn{1}{>{\raggedright}m{\dimexpr 1.54in+0\tabcolsep}}{\textcolor[HTML]{000000}{\fontsize{11}{11}\selectfont{variation}}} & \multicolumn{1}{>{\raggedleft}m{\dimexpr 1.35in+0\tabcolsep}}{\textcolor[HTML]{000000}{\fontsize{11}{11}\selectfont{5}}} \\

\multicolumn{1}{>{\raggedright}m{\dimexpr 1.52in+0\tabcolsep}}{\textcolor[HTML]{000000}{\fontsize{11}{11}\selectfont{ipilimumab}}} & \multicolumn{1}{>{\raggedright}m{\dimexpr 1.54in+0\tabcolsep}}{\textcolor[HTML]{000000}{\fontsize{11}{11}\selectfont{variation}}} & \multicolumn{1}{>{\raggedleft}m{\dimexpr 1.35in+0\tabcolsep}}{\textcolor[HTML]{000000}{\fontsize{11}{11}\selectfont{2}}} \\

\multicolumn{1}{>{\raggedright}m{\dimexpr 1.52in+0\tabcolsep}}{\textcolor[HTML]{000000}{\fontsize{11}{11}\selectfont{nivolumab}}} & \multicolumn{1}{>{\raggedright}m{\dimexpr 1.54in+0\tabcolsep}}{\textcolor[HTML]{000000}{\fontsize{11}{11}\selectfont{variation}}} & \multicolumn{1}{>{\raggedleft}m{\dimexpr 1.35in+0\tabcolsep}}{\textcolor[HTML]{000000}{\fontsize{11}{11}\selectfont{2}}} \\

\multicolumn{1}{>{\raggedright}m{\dimexpr 1.52in+0\tabcolsep}}{\textcolor[HTML]{000000}{\fontsize{11}{11}\selectfont{baloxavir\ marboxil}}} & \multicolumn{1}{>{\raggedright}m{\dimexpr 1.54in+0\tabcolsep}}{\textcolor[HTML]{000000}{\fontsize{11}{11}\selectfont{initial}}} & \multicolumn{1}{>{\raggedleft}m{\dimexpr 1.35in+0\tabcolsep}}{\textcolor[HTML]{000000}{\fontsize{11}{11}\selectfont{1}}} \\

\multicolumn{1}{>{\raggedright}m{\dimexpr 1.52in+0\tabcolsep}}{\textcolor[HTML]{000000}{\fontsize{11}{11}\selectfont{cemiplimab}}} & \multicolumn{1}{>{\raggedright}m{\dimexpr 1.54in+0\tabcolsep}}{\textcolor[HTML]{000000}{\fontsize{11}{11}\selectfont{variation}}} & \multicolumn{1}{>{\raggedleft}m{\dimexpr 1.35in+0\tabcolsep}}{\textcolor[HTML]{000000}{\fontsize{11}{11}\selectfont{1}}} \\

\multicolumn{1}{>{\raggedright}m{\dimexpr 1.52in+0\tabcolsep}}{\textcolor[HTML]{000000}{\fontsize{11}{11}\selectfont{dexmedetomidine}}} & \multicolumn{1}{>{\raggedright}m{\dimexpr 1.54in+0\tabcolsep}}{\textcolor[HTML]{000000}{\fontsize{11}{11}\selectfont{initial}}} & \multicolumn{1}{>{\raggedleft}m{\dimexpr 1.35in+0\tabcolsep}}{\textcolor[HTML]{000000}{\fontsize{11}{11}\selectfont{1}}} \\

\multicolumn{1}{>{\raggedright}m{\dimexpr 1.52in+0\tabcolsep}}{\textcolor[HTML]{000000}{\fontsize{11}{11}\selectfont{durvalumab}}} & \multicolumn{1}{>{\raggedright}m{\dimexpr 1.54in+0\tabcolsep}}{\textcolor[HTML]{000000}{\fontsize{11}{11}\selectfont{variation}}} & \multicolumn{1}{>{\raggedleft}m{\dimexpr 1.35in+0\tabcolsep}}{\textcolor[HTML]{000000}{\fontsize{11}{11}\selectfont{1}}} \\

\multicolumn{1}{>{\raggedright}m{\dimexpr 1.52in+0\tabcolsep}}{\textcolor[HTML]{000000}{\fontsize{11}{11}\selectfont{etirinotecan\ pegol}}} & \multicolumn{1}{>{\raggedright}m{\dimexpr 1.54in+0\tabcolsep}}{\textcolor[HTML]{000000}{\fontsize{11}{11}\selectfont{initial}}} & \multicolumn{1}{>{\raggedleft}m{\dimexpr 1.35in+0\tabcolsep}}{\textcolor[HTML]{000000}{\fontsize{11}{11}\selectfont{1}}} \\

\multicolumn{1}{>{\raggedright}m{\dimexpr 1.52in+0\tabcolsep}}{\textcolor[HTML]{000000}{\fontsize{11}{11}\selectfont{prasugrel}}} & \multicolumn{1}{>{\raggedright}m{\dimexpr 1.54in+0\tabcolsep}}{\textcolor[HTML]{000000}{\fontsize{11}{11}\selectfont{initial}}} & \multicolumn{1}{>{\raggedleft}m{\dimexpr 1.35in+0\tabcolsep}}{\textcolor[HTML]{000000}{\fontsize{11}{11}\selectfont{1}}} \\

\multicolumn{1}{>{\raggedright}m{\dimexpr 1.52in+0\tabcolsep}}{\textcolor[HTML]{000000}{\fontsize{11}{11}\selectfont{teriflunomide}}} & \multicolumn{1}{>{\raggedright}m{\dimexpr 1.54in+0\tabcolsep}}{\textcolor[HTML]{000000}{\fontsize{11}{11}\selectfont{variation}}} & \multicolumn{1}{>{\raggedleft}m{\dimexpr 1.35in+0\tabcolsep}}{\textcolor[HTML]{000000}{\fontsize{11}{11}\selectfont{1}}} \\

\ascline{1.5pt}{666666}{1-3}

\end{longtable}

\arrayrulecolor[HTML]{000000}

\global\setlength{\arrayrulewidth}{\Oldarrayrulewidth}

\global\setlength{\tabcolsep}{\Oldtabcolsep}

\renewcommand*{\arraystretch}{1}

\hypertarget{nph-issues-and-methods-to-address-them}{%
\section{NPH Issues and methods to address
them}\label{nph-issues-and-methods-to-address-them}}

We asked reviewers to determine whether in the reported assessment of
individual studies NPH was considered based no domain knowledge
(e.g.~known subgroup effects, delayed treatment effects with
immunotherapies), or post-hoc due to noteworthy results from model
diagnostics (e.g.~a statistically significant treatment by time
interaction). For more than half of the studies NPH was considered based
on domain knowledge (n=11). In the remaining studies either some model
diagnostic suggested a potential deviation from proportional hazards
(n=3), or the reason could not be determined (n=4). Table 3 provides the
number of studies per category. Please, notice that here the
observational unit is the number of studies (and not procedures) taking
into account that different reasons may have applied to different
studies.

\global\setlength{\Oldarrayrulewidth}{\arrayrulewidth}

\global\setlength{\Oldtabcolsep}{\tabcolsep}

\setlength{\tabcolsep}{0pt}

\renewcommand*{\arraystretch}{1.5}

\providecommand{\ascline}[3]{\noalign{\global\arrayrulewidth #1}\arrayrulecolor[HTML]{#2}\cline{#3}}

\begin{longtable}[c]{|p{2.06in}|p{1.07in}}

\caption{Number\ of\ trials\ for\ which\ NPH\ was\ either\ anticipated\ due\ to\ domain\ knowledge,\ or\ detected\ based\ no\ model\ diagnostics.}\\

\ascline{1.5pt}{666666}{1-2}

\multicolumn{1}{>{\raggedright}m{\dimexpr 2.06in+0\tabcolsep}}{\textcolor[HTML]{000000}{\fontsize{11}{11}\selectfont{Why\ was\ NPH\ considered}}} & \multicolumn{1}{>{\raggedleft}m{\dimexpr 1.07in+0\tabcolsep}}{\textcolor[HTML]{000000}{\fontsize{11}{11}\selectfont{\#\ of\ studies}}} \\

\ascline{1.5pt}{666666}{1-2}\endhead

\multicolumn{1}{>{\raggedright}m{\dimexpr 2.06in+0\tabcolsep}}{\textcolor[HTML]{000000}{\fontsize{11}{11}\selectfont{domain\ knowledge}}} & \multicolumn{1}{>{\raggedleft}m{\dimexpr 1.07in+0\tabcolsep}}{\textcolor[HTML]{000000}{\fontsize{11}{11}\selectfont{11}}} \\

\multicolumn{1}{>{\raggedright}m{\dimexpr 2.06in+0\tabcolsep}}{\textcolor[HTML]{000000}{\fontsize{11}{11}\selectfont{model\ diagnostics}}} & \multicolumn{1}{>{\raggedleft}m{\dimexpr 1.07in+0\tabcolsep}}{\textcolor[HTML]{000000}{\fontsize{11}{11}\selectfont{3}}} \\

\multicolumn{1}{>{\raggedright}m{\dimexpr 2.06in+0\tabcolsep}}{\textcolor[HTML]{000000}{\fontsize{11}{11}\selectfont{unclear}}} & \multicolumn{1}{>{\raggedleft}m{\dimexpr 1.07in+0\tabcolsep}}{\textcolor[HTML]{000000}{\fontsize{11}{11}\selectfont{4}}} \\

\ascline{1.5pt}{666666}{1-2}

\end{longtable}

\arrayrulecolor[HTML]{000000}

\global\setlength{\arrayrulewidth}{\Oldarrayrulewidth}

\global\setlength{\tabcolsep}{\Oldtabcolsep}

\renewcommand*{\arraystretch}{1}

\hypertarget{suspected-causes-of-nph}{%
\subsubsection{Suspected causes of NPH}\label{suspected-causes-of-nph}}

Reviewers were asked to assess the suspected cause for NPH based on the
discussion in the EPAR. Corresponding answers were categorized into
broad groups of potential causes. Table 4 reports the number of studies
were a specific cause for NPH was suspected. For most studies the
suspected cause for NPH was a `delayed treatment effect'. This was
sometimes considered in sample size planning by assuming different
hazard ratios for earlier and later time-periods. Also differential
treatment effects were anticipated for a large proportion of studies
(e.g.~trials of pembrolizumab, nivolumab). This was typically addressed
by stratification, both in terms of randomization and analysis method
(e.g.~stratified log-rank test). Mostly, corresponding EPARs reported
extensive subgroup analyses evaluating the differential treatment effect
in the respective subgroups. Other potential causes were related to
censoring, or treatment switching. Especially for the case of
\emph{teriflunimide} - a treatment for multiple sclerosis, for which
authorization was to be extended to a pediatric population based on a
trial with primary endpoint time to clinical relapse - the intercurrent
event of rescue following high MRI activity was considered the potential
cause for a waning treatment effect. In this case the issue was
addressed by additional analyses targeting a composite estimand, where
the use of rescue was counted as clinical relapse in a corresponding
composite endpoint. For \emph{baloxavir marboxil} - a treatment for
human influenza based on two trials of the compound with an endpoint of
time to recovery from influenza infection - NPH was anticipated due to
the assumption that all subjects recover within the follow-up period,
which would be incompatible with an assumption of proportional hazards.

\global\setlength{\Oldarrayrulewidth}{\arrayrulewidth}

\global\setlength{\Oldtabcolsep}{\tabcolsep}

\setlength{\tabcolsep}{0pt}

\renewcommand*{\arraystretch}{1.5}

\providecommand{\ascline}[3]{\noalign{\global\arrayrulewidth #1}\arrayrulecolor[HTML]{#2}\cline{#3}}

\begin{longtable}[c]{|p{5.22in}|p{1.07in}}

\caption{Number\ of\ suspected\ causes\ for\ NPH\ in\ trials\ included\ in\ the\ review.}\\

\ascline{1.5pt}{666666}{1-2}

\multicolumn{1}{>{\raggedright}m{\dimexpr 5.22in+0\tabcolsep}}{\textcolor[HTML]{000000}{\fontsize{11}{11}\selectfont{Suspected\ cause\ for\ NPH}}} & \multicolumn{1}{>{\raggedleft}m{\dimexpr 1.07in+0\tabcolsep}}{\textcolor[HTML]{000000}{\fontsize{11}{11}\selectfont{\#\ of\ studies}}} \\

\ascline{1.5pt}{666666}{1-2}\endhead

\multicolumn{1}{>{\raggedright}m{\dimexpr 5.22in+0\tabcolsep}}{\textcolor[HTML]{000000}{\fontsize{11}{11}\selectfont{delayed\ treatment\ effect}}} & \multicolumn{1}{>{\raggedleft}m{\dimexpr 1.07in+0\tabcolsep}}{\textcolor[HTML]{000000}{\fontsize{11}{11}\selectfont{3}}} \\

\multicolumn{1}{>{\raggedright}m{\dimexpr 5.22in+0\tabcolsep}}{\textcolor[HTML]{000000}{\fontsize{11}{11}\selectfont{delayed\ treatment\ effect,\ censoring}}} & \multicolumn{1}{>{\raggedleft}m{\dimexpr 1.07in+0\tabcolsep}}{\textcolor[HTML]{000000}{\fontsize{11}{11}\selectfont{1}}} \\

\multicolumn{1}{>{\raggedright}m{\dimexpr 5.22in+0\tabcolsep}}{\textcolor[HTML]{000000}{\fontsize{11}{11}\selectfont{delayed\ treatment\ effect,\ differential\ treatment\ effect}}} & \multicolumn{1}{>{\raggedleft}m{\dimexpr 1.07in+0\tabcolsep}}{\textcolor[HTML]{000000}{\fontsize{11}{11}\selectfont{3}}} \\

\multicolumn{1}{>{\raggedright}m{\dimexpr 5.22in+0\tabcolsep}}{\textcolor[HTML]{000000}{\fontsize{11}{11}\selectfont{delayed\ treatment\ effect,\ treatment\ switching,\ differential\ treatment\ effect}}} & \multicolumn{1}{>{\raggedleft}m{\dimexpr 1.07in+0\tabcolsep}}{\textcolor[HTML]{000000}{\fontsize{11}{11}\selectfont{1}}} \\

\multicolumn{1}{>{\raggedright}m{\dimexpr 5.22in+0\tabcolsep}}{\textcolor[HTML]{000000}{\fontsize{11}{11}\selectfont{differential\ treatment\ effect}}} & \multicolumn{1}{>{\raggedleft}m{\dimexpr 1.07in+0\tabcolsep}}{\textcolor[HTML]{000000}{\fontsize{11}{11}\selectfont{3}}} \\

\multicolumn{1}{>{\raggedright}m{\dimexpr 5.22in+0\tabcolsep}}{\textcolor[HTML]{000000}{\fontsize{11}{11}\selectfont{unclear}}} & \multicolumn{1}{>{\raggedleft}m{\dimexpr 1.07in+0\tabcolsep}}{\textcolor[HTML]{000000}{\fontsize{11}{11}\selectfont{7}}} \\

\ascline{1.5pt}{666666}{1-2}

\end{longtable}

\arrayrulecolor[HTML]{000000}

\global\setlength{\arrayrulewidth}{\Oldarrayrulewidth}

\global\setlength{\tabcolsep}{\Oldtabcolsep}

\renewcommand*{\arraystretch}{1}

\hypertarget{primary-analysis-methods}{%
\subsubsection{Primary analysis
methods}\label{primary-analysis-methods}}

For most of the trials included in the review the primary analysis
method was either a (stratified) log-rank test or (stratified) CoxPH
model. As discussed above, for studies where NPH was anticipated due to
a delayed treatment effect, NPH was often addressed by planning a larger
sample size to obtain sufficient power using conventional PH methods.
Some notable exceptions were the trial of \emph{prasugrel} - a treatment
for the prevention of atherothrombotic events in patients with acute
coronary syndrome. In this trial the primary analysis was a
Gehan-Wilcoxon test comparing the time to event for the composite
endpoint of cardiovascular death, nonfatal myocardial infarction, or
nonfatal stroke. For \emph{dexmedetomidine} - a treatment for patients
requiring light to moderate sedation in intensive care during or after
intubation - the primary analysis to compare the duration of mechanical
ventilation was either a CoxPH or Gehan-Wilcoxon test depending on
whether a test for a treatment by time interaction was significant. For
\emph{baloxavir marboxil} the primary analysis of the endpoint time to
recovery from influenza infection was a stratified Peto-Prentice
generalized Wilcoxon model. Table 5 reports the number of times a
specific analysis method was reported as primary or sensitivity analysis
in on of the EPARs.

\global\setlength{\Oldarrayrulewidth}{\arrayrulewidth}

\global\setlength{\Oldtabcolsep}{\tabcolsep}

\setlength{\tabcolsep}{0pt}

\renewcommand*{\arraystretch}{1.5}

\providecommand{\ascline}[3]{\noalign{\global\arrayrulewidth #1}\arrayrulecolor[HTML]{#2}\cline{#3}}

\begin{longtable}[c]{|p{3.32in}|p{1.07in}}

\caption{Number\ and\ types\ of\ statistical\ methods\ used\ for\ analysis\ (both\ primary\ and\ sensitivity\ analyses)}\\

\ascline{1.5pt}{666666}{1-2}

\multicolumn{1}{>{\raggedright}m{\dimexpr 3.32in+0\tabcolsep}}{\textcolor[HTML]{000000}{\fontsize{11}{11}\selectfont{Method}}} & \multicolumn{1}{>{\raggedleft}m{\dimexpr 1.07in+0\tabcolsep}}{\textcolor[HTML]{000000}{\fontsize{11}{11}\selectfont{\#\ of\ studies}}} \\

\ascline{1.5pt}{666666}{1-2}\endhead

\multicolumn{1}{>{\raggedright}m{\dimexpr 3.32in+0\tabcolsep}}{\textcolor[HTML]{000000}{\fontsize{11}{11}\selectfont{Logrank\ test}}} & \multicolumn{1}{>{\raggedleft}m{\dimexpr 1.07in+0\tabcolsep}}{\textcolor[HTML]{000000}{\fontsize{11}{11}\selectfont{16}}} \\

\multicolumn{1}{>{\raggedright}m{\dimexpr 3.32in+0\tabcolsep}}{\textcolor[HTML]{000000}{\fontsize{11}{11}\selectfont{CoxPH\ model}}} & \multicolumn{1}{>{\raggedleft}m{\dimexpr 1.07in+0\tabcolsep}}{\textcolor[HTML]{000000}{\fontsize{11}{11}\selectfont{14}}} \\

\multicolumn{1}{>{\raggedright}m{\dimexpr 3.32in+0\tabcolsep}}{\textcolor[HTML]{000000}{\fontsize{11}{11}\selectfont{RMST}}} & \multicolumn{1}{>{\raggedleft}m{\dimexpr 1.07in+0\tabcolsep}}{\textcolor[HTML]{000000}{\fontsize{11}{11}\selectfont{3}}} \\

\multicolumn{1}{>{\raggedright}m{\dimexpr 3.32in+0\tabcolsep}}{\textcolor[HTML]{000000}{\fontsize{11}{11}\selectfont{Wilcoxon\ type\ test}}} & \multicolumn{1}{>{\raggedleft}m{\dimexpr 1.07in+0\tabcolsep}}{\textcolor[HTML]{000000}{\fontsize{11}{11}\selectfont{5}}} \\

\multicolumn{1}{>{\raggedright}m{\dimexpr 3.32in+0\tabcolsep}}{\textcolor[HTML]{000000}{\fontsize{11}{11}\selectfont{Rank\ preserving\ structural\ failure\ time\ model}}} & \multicolumn{1}{>{\raggedleft}m{\dimexpr 1.07in+0\tabcolsep}}{\textcolor[HTML]{000000}{\fontsize{11}{11}\selectfont{2}}} \\

\ascline{1.5pt}{666666}{1-2}

\end{longtable}

\arrayrulecolor[HTML]{000000}

\global\setlength{\arrayrulewidth}{\Oldarrayrulewidth}

\global\setlength{\tabcolsep}{\Oldtabcolsep}

\renewcommand*{\arraystretch}{1}

\hypertarget{sensitivity-analyses-and-model-diagnostics}{%
\subsubsection{Sensitivity analyses and model
diagnostics}\label{sensitivity-analyses-and-model-diagnostics}}

For many of the included studies, discussion of NPH issues was limited
to model diagnostics and sensitivity analyses. The most common model
diagnostic used to detect potential NPH were tests for a treatment by
time interaction in a proportional hazards model. Other notable methods
used were visual inspection of the survival curves, visual inspection of
the Schönfeld residuals and log-log plots. For some trials estimates of
the RMST in different time-periods, piece-wise hazard ratios, kernel
based estimates of the log hazards, or results from structural failure
time models were provided.

\hypertarget{extracted-study-characteristics-and-results}{%
\section{Extracted study characteristics and
results}\label{extracted-study-characteristics-and-results}}

Reviewers were asked to extract data on the primary analysis results,
including effects tables, subgroup analyses, as well as, results from
sensitivity and supportive analyses addressing NPH issues. Where not
available, important outcomes were extracted after full-text review from
EPARs. For some procedures, where particular results were not reported
in the EPAR, corresponding results were obtained from digitized survival
curves. Survival curves for the primary endpoint were digitized for the
majority of trials with a focus on trials with delayed treatment
effects.

\hypertarget{sample-size}{%
\subsubsection{Sample size}\label{sample-size}}

For all trials we extracted the number of subjects included in the
experimental group. For trials investigating more than one experimental
treatment arm we selected the arm studying a higher dose, or the largest
combination (e.g.~where the experimental treatment was studied as a
mono-therapy and combination treatment). Table 6 reports the average
number of subjects in the treatment group for trials within each
indication area. In the oncology domain trials recruited an average of
about 400 subjects to the treatment group, with sample numbers ranging
from about 150 to about 800. In all other indications only one procedure
was identified, however with two trials reported for each of the
procedures in \emph{influenza (human)} and \emph{conscious sedation},
respectively. Sample numbers range from about 100 for the pediatric
trial in \emph{multiple sclerosis} to about 7000 for the trial in the
\emph{cardiovascular disease} area.

\global\setlength{\Oldarrayrulewidth}{\arrayrulewidth}

\global\setlength{\Oldtabcolsep}{\tabcolsep}

\setlength{\tabcolsep}{0pt}

\renewcommand*{\arraystretch}{1.5}

\providecommand{\ascline}[3]{\noalign{\global\arrayrulewidth #1}\arrayrulecolor[HTML]{#2}\cline{#3}}

\begin{table}[ht]
\centering
\begin{tabular}{p{1.4in}>{\raggedleft\arraybackslash}p{1.09in}>{\raggedleft\arraybackslash}p{2.7in}>{\raggedleft\arraybackslash}p{0.67in}>{\raggedleft\arraybackslash}p{0.67in}}
\hline
\textbf{Indication Area} & \textbf{\# of Studies} & \textbf{Avg. Sample Size Treatment Group} & \textbf{Min} & \textbf{Max} \\
\hline
Oncology & 12 & 387 & 153 & 789 \\
Conscious sedation & 2 & 250 & 249 & 251 \\
Influenza (human) & 2 & 671 & 612 & 730 \\
Cardiovascular disease & 1 & 6813 & 6813 & 6813 \\
Multiple sclerosis & 1 & 109 & 109 & 109 \\
\hline
\end{tabular}
\caption{Average, minimum and maximum sample size across trials per indication area.}
\end{table}
\arrayrulecolor[HTML]{000000}

\global\setlength{\arrayrulewidth}{\Oldarrayrulewidth}

\global\setlength{\tabcolsep}{\Oldtabcolsep}

\renewcommand*{\arraystretch}{1}

\hypertarget{subject-recruitment-and-study-duration}{%
\subsubsection{Subject recruitment and study
duration}\label{subject-recruitment-and-study-duration}}

\begin{figure}
\centering
\includegraphics{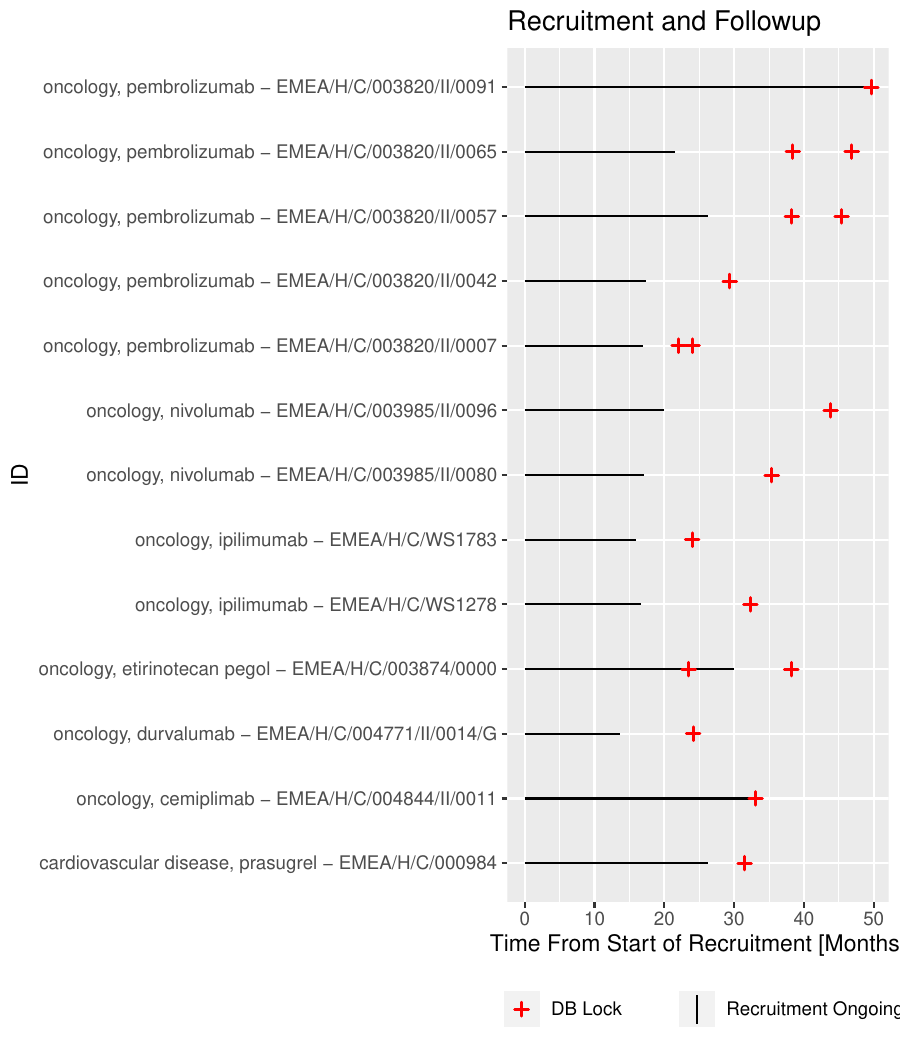}
\caption{Follow-up and recruitment times for selected studies
in months. Solid line indicates months with ongoing recruitment from
study initiation. Red crosses indicate times of data-base lock. Multiple
data-base lock times are shown in case of interim analyses.}\label{fig:accrual}
\end{figure}

We extracted the duration of study duration and time from study
initiation to data-base lock for included studies. Figure \ref{fig:accrual} presents
corresponding results for all studies except those related to procedures
3788 and 9999 for which corresponding data could not be found. On
average studies recruited subjects over a period of 23.4154232 months.
The average maximum follow-up at the time of confirmatory analysis was
35.2103265 months.

\hypertarget{median-survival}{%
\subsubsection{Median survival}\label{median-survival}}

\begin{figure}
\centering
\includegraphics{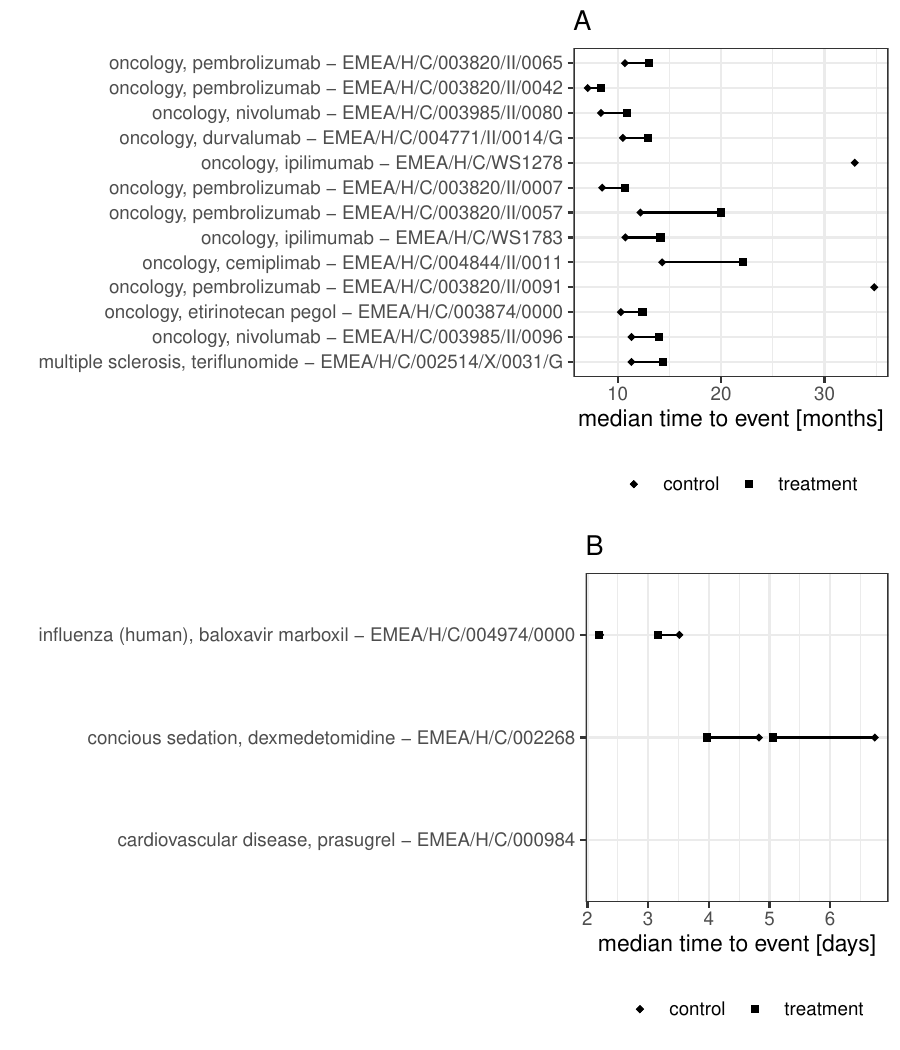}
\caption{Effect size estimates from EPARs. Squares indicate
median time to event in the control group, circles median time to event
in the treatment group. For procedures where either one or both
estimates are missing, follow-up was insufficient to estimate respective
survival quantiles. A: trials with time to event measured in months. B:
trials with time to event measured in days.}\label{fig:mediansurvival}
\end{figure}

We extracted the median survival with respect to the primary enpdoint in
the experimental and control arm for all of the trials included. For
trials investigating more than one experimental treatment arm we
selected the arm studying a higher dose, or the largest combination
(e.g.~where the experimental treatment was studied as a mono-therapy and
combination treatment). Figure \ref{fig:mediansurvival} presents the extracted median survival
times for the included studies. Figure \ref{fig:mediansurvival}A provides results for trials
where the time to event was studied on a scale of months, whereas Figure
\ref{fig:mediansurvival}B provides results for trials where the time to event was studied on a
scale of days. The former all represent trials, where the event of
interest corresponds to a negative outcome (e.g.~time to all-cause
mortality, time to clinical relapse). For the trials of \emph{baloxavir
marboxil} and \emph{dexmedetomidine} the event represents a positive
outcome (time to recovery, duration of mechanical ventilation). For two
of the oncology trials, \emph{ipilimumab - EMEA/H/C/WS1278}, and
\emph{pembrolizumab - EMEA/H/C/003820/II/0091}, the median time to event
was not reached in the treatment groups. Similar for the trial of
\emph{prasugrel - EMEA/H/C/000984}, follow-up time was not sufficient to
estimate the median time to event in either treatment group.

\hypertarget{delayed-treatment-effect}{%
\subsubsection{Delayed treatment
effect}\label{delayed-treatment-effect}}

For procedures where a delayed treatment effect was suspected, we
extracted the time of survival curve separation (i.e.~the timepoint
after which a benefit of the experimental treatment can be observed in
the KM plots) from EPARs. Where available, we extracted the estimated
time of separation from the efficacy discussion in the EPAR. In
addition, we digitized the KM plots from EPARs (using R Package
\texttt{IPDfromKM} - \citep{Liu2021}) and adjudicated the time of
separation based on visual inspection. Table 7 provides the times of
separation (both extracted and adjudicated) for included procedures
where a delayed treatment effect was suspected. All procedures refer to
monoclonal antibodies from the oncology domain. The adjudicated time of
separation ranges from about 2 months to up to about 9 months. While for
some procedures the KM curves overlap during the first period of
treatment - with no clear benefit for either treatment during an earlier
treatment period (e.g.~\emph{pembrolizumab - EMEA/H/C/003820/II/0065}),
for others the KM curves cross - with a pronounced survival benefit
during a first period of treatment which is reversed to a survival
benefit in the experimental group later on (e.g.~\emph{nivolumab -
EMEA/H/C/003985/II/0080}).

\global\setlength{\Oldarrayrulewidth}{\arrayrulewidth}

\global\setlength{\Oldtabcolsep}{\tabcolsep}

\setlength{\tabcolsep}{0pt}

\renewcommand*{\arraystretch}{1.5}

\providecommand{\ascline}[3]{\noalign{\global\arrayrulewidth #1}\arrayrulecolor[HTML]{#2}\cline{#3}}

\begin{table}[ht]
\centering
\caption{Time of survival curve separation for trials with delayed treatment effects. Column three provides the time of delay/separation as reported in the EPAR, whereas the last column provides the time of separation adjudicated from digitized KM curves.}
\begin{tabular}{p{1.47in}p{1.5in}>{\raggedleft\arraybackslash}p{1.5in}>{\raggedleft\arraybackslash}p{1.5in}}
\hline
\textbf{Active Substance} & \textbf{Procedure Number} & \textbf{Rep. Separation} & \textbf{Adj. Separation} \\
\hline
nivolumab & EMEA/H/C/003985/II/0096 & 6 & 1.7 \\
ipilimumab & EMEA/H/C/WS1783 &  & 3.6 \\
ipilimumab & EMEA/H/C/WS1278 &  & 3.8 \\
cemiplimab & EMEA/H/C/004844/II/0011 & 6 & 4.0 \\
nivolumab & EMEA/H/C/003985/II/0080 & 5 & 4.8 \\
pembrolizumab & EMEA/H/C/003820/II/0042 & 5 & 4.8 \\
pembrolizumab & EMEA/H/C/003820/II/0065 & 8 & 8.5 \\
pembrolizumab & EMEA/H/C/003820/II/0091 & 8 & 8.7 \\
\hline
\end{tabular}
\end{table}

\arrayrulecolor[HTML]{000000}

\global\setlength{\arrayrulewidth}{\Oldarrayrulewidth}

\global\setlength{\tabcolsep}{\Oldtabcolsep}

\renewcommand*{\arraystretch}{1}

\hypertarget{differential-treatment-effects-in-subgroups}{%
\subsubsection{Differential treatment effects in
subgroups}\label{differential-treatment-effects-in-subgroups}}

\begin{figure}
\includegraphics[width=6in]{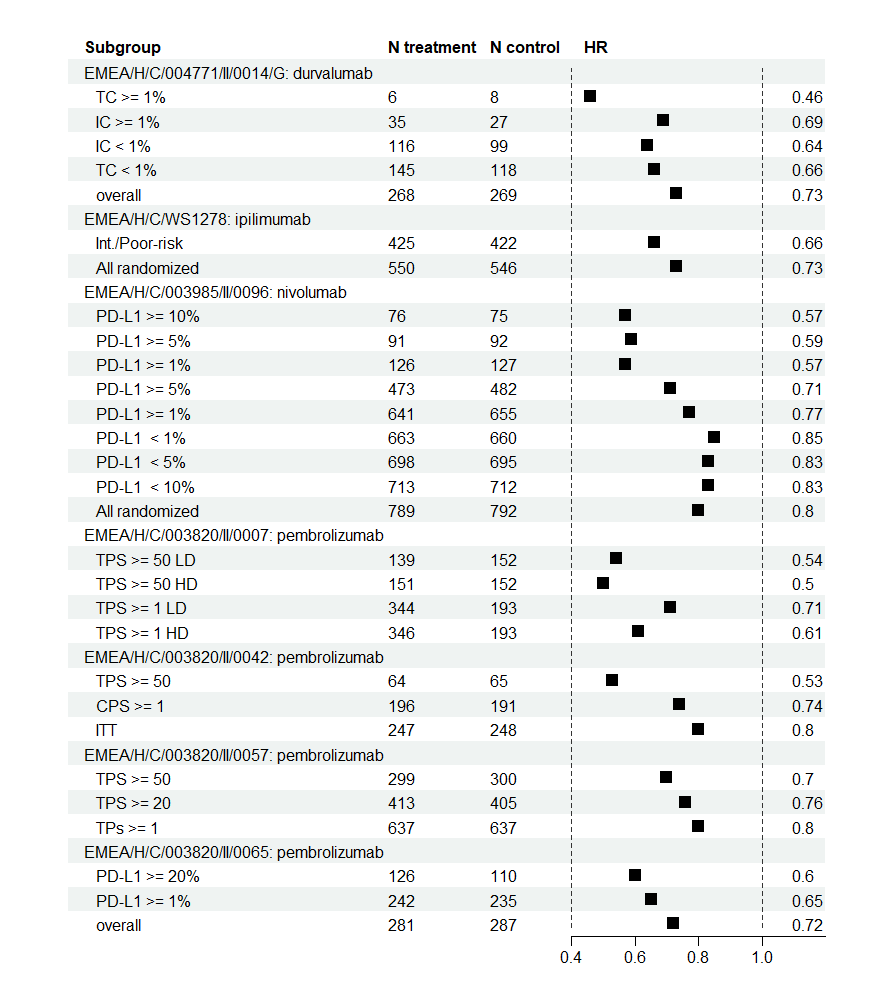} \caption{Forest Plot for selected studies with differential treatment effect}\label{fig:forest}
\end{figure}

For a majority of the products included in the review, a differential
treatment effect in defined subgroups was expected. Especially, this
applied to the monoclonal antibodies for cancer therapy. In the analysis
of these trials the resulting violation of the proportional hazards
assumption was addressed by using stratified analyses. Often extensive
subgroup analyses were reported. Figure \ref{fig:forest} presents estimates of the
hazard ratio for various various biomarker defined subgroups, which were
extracted from the respective EPARs. Depending on the trial the
proportion of the corresponding subgroups ranged from 10\% to more than
50\%. Differences in hazard ratios between overall population and
subgroups were substantial (e.g.~\emph{pembrolizumab -
EMEA/H/C/003820/II/0065} HR=0.72 overall, HR=0.6 for subjects with PD-L1
\textgreater= 20\%). It should be noted, however, that subgroups were
mostly defined based on a pre-defined cut-off for some continuous
biomarker (e.g.~PD-L1 \textgreater= 10\%) and it is unclear whether
efficacy is not in fact continuously related to this marker.

\hypertarget{limitations}{%
\section{Limitations}\label{limitations}}

While we were able to identify a substantial number of procedures where
issues with NPH were identified in the risk-benefit assessment, the
present review should not be misunderstood as a representative or even
complete sample of marketing authorization procedures with NPH issues.
In line with the objectives of the review, we only included procedures
where a substantial discussion of NPH issues was reflected in the EPAR.

Considering that conventional methods like the log-rank test do control
the Type I error rate even under NPH, one can assume that violations of
the proportional hazards assumption may often be overlooked or given
secondary priority in the assessment. This would explain our finding
that methods addressing NPH are rarely planned as the primary analysis
in pivotal trials. Furthermore, model diagnostics (e.g.~testing for a
treatment by time interaction) may have poor sensitivity as the trials
are typically not powered to detect corresponding effects. Consequently,
there may be a number procedures, where the proportional hazards
assumption did not apply but the issue was not detected during
assessment. Corresponding procedures would not have matched our search
and eligibility criteria and are, therefore, not represented in the
review.

Moreover, the methodological aspects of the risk-benefit assessment are
not always reflected in full in the EPAR. Consequently, there may be a
number of procedure, where NPH might have been identified, but the
corresponding discussion not reproduced in the EPAR. Corresponding
procedures would not have matched our search and eligibility criteria
and are, therefore, not represented in the review. Similarly, the need
to constrain the methodological discussion in EPARs to the most
essential findings, may limit the amount of results reported for
e.g.~sensitivity analyses and model diagnostics available. This has been
noticed in the review, as for quite some procedures statistical methods
addressing NPH were mentioned, but results not reported.

Except for one procedure - \emph{etirinotecan pegol -
EMEA/H/C/003874/0000} - referred to us by EMA and identified using the
search engine at \url{https://www.ema.europa.eu/en/medicines}, we also
excluded procedures with a negative opinion. This decision was primarily
driven by limitations of the database at \url{https://paediatricdata.eu}
which may not comprehensively cover withdrawn applications.

\hypertarget{summary-and-conclusion}{%
\section{Summary and conclusion}\label{summary-and-conclusion}}

Despite all limitations discussed above, our review of past marketing
authorizations identified 16 procedures reporting results from 18
distinct trials, where non-proportional hazards were identified as an
issue in the risk-benefit assessment . The majority of the corresponding
treatments are from the oncology domain with exceptions from influenza
(human), multiple sclerosis, conscious sedation, and cardiovascular
disease. For the majority of procedures from the oncology domain,
non-proportional hazards were anticipated to be related to a delayed,
and/or a differential treatment effect. This can be explained by the
large number of procedures covering immunotherapies in oncology.
However, also issues corresponding to censoring or treatment switching
were noted for some procedures.

Even though, non-proportional hazards were anticipated in a large
proportion of trials, the primary analysis was still performed using
conventional methods assuming proportional hazards. While, differential
treatment effects were addressed using stratification. Delayed treatment
effects were not addressed in the analysis model, but rather by
recruiting a larger sample size to compensate for the related loss in
power. The, at times, substantial discussion of NPH issues in EPARs was
mostly limited to various subgroup analyses, sensitivity analyses, and
model diagnostics. However, in a few cases weighted log-rank tests were
planned for the primary analysis, sometimes conditional on a pre-test of
the proportional hazards assumption.

We managed to extract a wealth of results from the included studies.
This includes: sample sizes, median survival times, hazard ratios. In
addition, we digitized survival curves, allowing us to reconstruct
specific examples close to individual patient data. Consequently, we
obtained a wide range of data from actual trials, that can be used to
determine relevant parameter ranges for the planning of simulation
studies to evaluate the performance characteristics of alternative
analysis methods. Especially, reconstructed individual patient level
data will be useful to construct case studies to illustrate individual
findings, based on re-analyses of the data. For example \citep{Klinglmuller2023} report results from an extensive simulation study. Underlying
assumptions and parameter ranges were based on results from this review
and prospectively specified. The study protocol was pre-registered and
can be found at \url{https://catalogues.ema.europa.eu/node/3500}.
Moreover, the various types of NPH issues identified and the related
discussions in the identified EPARs will be useful to derive regulatory
recommendations on how to address the corresponding implications in
future marketing authorization procedures.

Finally, we note that although the number of procedures identified is
substantial. The majority of procedures cover relatively recent
immunotherapies from oncology. Specifically, checkpoint inhibitors with
know differential subgroup effects, and suspected delayed treatment
effects (e.g.~when compared to chemotherapy). Consideration, of NPH in
the assessment may be driven by the growing awareness in the scientific
community that conventional methods assuming proportional hazards may
not be optimal in these indications. The relative lack of procedures
identified from other areas, could indicate that the issue of
non-proportional hazards may not have received the attention in other
areas it potential deserves.

\hypertarget{funding}{%
\section{Funding}\label{funding}}

This work has received funding from the European Medicines Agency
(Re-opening of competition EMA/2020/46/TDA/L3.02 (Lot 3))

\bibliographystyle{plain}

\bibliography{references}

\hypertarget{appendix-i-table-of-searchterms-with-number-of-returned-paragraphs-per-term}{%
\section{Appendix I: Table of search terms with number of returned
paragraphs per
term}\label{appendix-i-table-of-searchterms-with-number-of-returned-paragraphs-per-term}}

\begin{figure}
\includegraphics[width=6in]{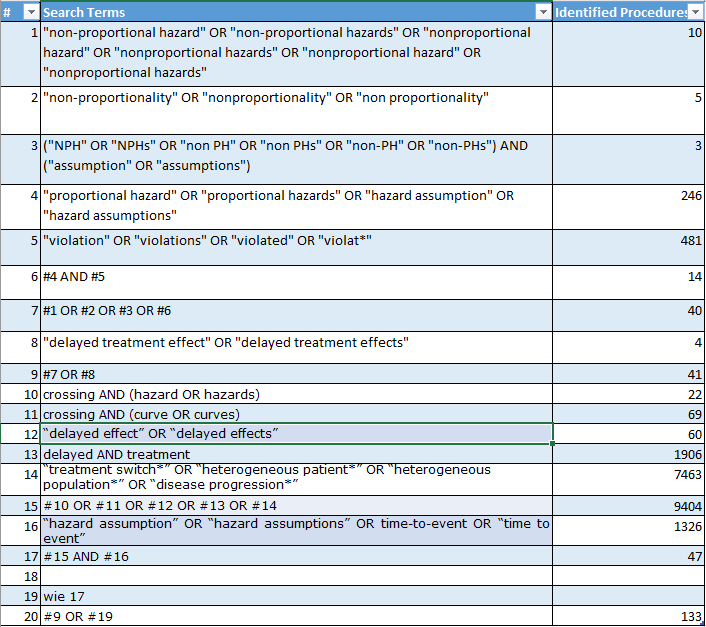} \caption{Search terms with detailed results}\label{fig:unnamed-chunk-5}
\end{figure}

\end{document}